\documentclass[aps,preprint]{revtex4}
\input{epsf}

\begin{document}

\def\be{\begin{equation}}
\def\ee{\end{equation}}
\def\bea{\begin{eqnarray}}
\def\eea{\end{eqnarray}}
\def\ba{\begin{array}}
\def\ea{\end{array}}


\title{Radion Stabilization in Compact Hyperbolic Extra Dimensions}

\author{Salah Nasri$^{1}$, Pedro J. Silva$^{1}$, Glenn D. Starkman$^{2}$ and Mark
Trodden$^{1}$}

\affiliation{~\\
$^{1}$ Physics Department, \\
Syracuse University, \\
Syracuse, New York 13244-1130. \\
\\
$^{2}$ Department of Physics, \\
Case Western Reserve University, \\
Cleveland, OH 44106-7079.}

\begin{abstract}
We consider radion stabilization in hyperbolic brane-world scenarios. We demonstrate that
in the context of Einstein gravity, matter fields which stabilize the extra dimensions
must violate the null energy condition. This result is shown to hold even allowing for FRW-like
expansion on the brane. In particular, we explicitly demonstrate how one
putative source of stabilizing matter fails to work, and how others violate the above condition.
We speculate on a number of ways in which we may bypass this result, including the effect of
Casimir energy in these spaces. A brief discussion of supersymmetry in these backgrounds is also
given.
\end{abstract}

\preprint{SU-GP-02/1-1}
\preprint{SU-4252-754}
\preprint{CWRU-01-02}

\maketitle

\section{Introduction}
Unification physics has traditionally been seen as the problem of reconciling
wildly disparate mass scales, for example the weak scale ($\sim 10^{2}$GeV) and
the Planck scale ($\sim 10^{19}$GeV). This exponential hierarchy is technically
unnatural in particle physics, since in general, the effects of renormalization
are to make the observable values of such scales much closer in size.
Well-known attempts to address this issue, such as supersymmetry (SUSY) in which
delicate cancellations between renormalization terms occur, or technicolor, in
which the renormalization effects are much less dramatic than one might ordinarily
expect, function by preventing dramatic corrections to an externally imposed
mass hierarchy.

A fresh perspective on the problem of unification has received much attention in recent
years \cite{Anton,ADD,RS,otherearly}. In
this picture the hierarchy problem is is no longer a disparity between mass scales,
and instead becomes an issue of length scales. The new approach is a
superstring-inspired modification of the Kaluza-Klein idea that the
universe may have more spatial dimensions than the three that we observe. The
general hypothesis is that the universe as a whole is $3+1+d$ dimensional, with
gravity propagating in all dimensions, but the standard model fields are confined
to a $3+1$ dimensional submanifold that comprises our observable universe. The
primary motivation for this comes from Polchinski's discovery \cite{polchinski}
of $D$-branes in string theory. These extended objects have the property that open
strings, the excitations of which correspond to standard model particles, may end
on them, and thus are confined to the brane. However, closed string excitations,
corresponding to gravitational degrees of freedom are free to occur anywhere in the
space.

As in traditional Kaluza-Klein theories, it is necessary that all dimensions other
than those we observe be compactified, so that their existence does not conflict
with experimental data. The difference in the new scenarios is that, since standard
model fields do not propagate in the extra dimensions, it is only necessary to
evade constraints on higher-dimensional gravity, and not, for example, on
higher-dimensional electromagnetism. As we shall see, this is important, since
electromagnetism is tested to great precision down to extremely small scales,
whereas microscopic tests of gravity are far less precise.

Since constraints on the new scenarios are less stringent than those on ordinary
Kaluza-Klein theories, the corresponding extra dimensions can be significantly
larger, which translates into a much larger allowed volume for the extra dimensions.
It is the spreading of gravitational flux into this large volume that allows
gravity measured on our $3$-brane to be so weak (parameterized by the Planck mass,
$M_P$), while the fundamental scale of physics $M_{d+4}$ is parameterized by the
weak scale, $M_W$ say. Thus, the problem of understanding the hierarchy between
the Planck and weak scales now becomes that of understanding why extra dimensions
are stabilized at a volume
large in units of the fundamental length scale $M_{d+4}^{-1}$.
This is the rephrasing of the
hierarchy problem in these models. It constitutes a fundamental shift in thinking.
Traditionally, these large compact  extra dimensions  have been conceived of
as d-tori, or d-spheres.  In this setting, one has the added bonus of requiring a
linear tuning of length scales, compared
to the usual exponential tuning of mass scales. Nevertheless, a significant tuning
is still required, although now in an entirely different sector of the theory.

In recent work \cite{Kaloper:2000jb,Trodden:2000cm} two of us proposed a modification to the
above picture,
in which we argued that there exist attractive alternate
choices of compactification.  These compactifications employ a topologically
non-trivial internal space -- a $d$-dimensional compact hyperbolic manifold (CHM).
They also throw into a new light the problem of explaining the large hierarchy
$M_P/{\rm TeV}$, since even though the volume of these manifolds is large,
their linear size $L$ is only slightly larger than the new fundamental
length scale ($L\sim 30 M_{d+4}^{-1}$ for example), thus only requiring numbers of
${\cal O}(10)$. Further, cosmology in such spaces has interesting consequences
for the evolution of the early universe \cite{Starkman:2001xu,Starkman:2001dy}.
In the next section we provide a brief review of the relevant properties of CHMs.

The main purpose of this paper is to present a detailed analysis of radion stabilization in
these models. It has recently been demonstrated \cite{Carroll:2001ih,otherrefs} that, in the
context of
general relativity in $4+d$ dimensions, stabilization of large hyperbolic extra dimensions,
leaving Minkowski space on our brane, requires a violation of the null dominant energy
condition. In section \ref{stabilization} we extend this argument to the case in which our
brane is allowed to exhibit standard FRW expansion, and comment on the regime of validity of this
result. We then turn to possible ways in which stabilization may work due to a breakdown of the
assumptions in the previous argument or through quantum stabilization effects. We provide an
explicit example of this possibility through the Casimir force in CHMs.

For completeness we include some final comments on supersymmetry in compact hyperbolic backgrounds
before concluding.

\section{Compact Hyperbolic Manifolds and Extra Dimensions}
A $d$-dimensional compact hyperbolic manifold has spatial sections of the form
$\Sigma= H^d/\Gamma$, where the fundamental group, $\Gamma$,
is a discrete subgroup of
$SO(d,1)$ acting freely ({\it ie.} without fixed
points) and discontinuously (since it is discrete). The CHM can be
obtained by gluing together the faces of a fundamental domain in hyperbolic
space.

Hyperbolic space in $d$ dimensions can be viewed as the hyperboloid
\begin{equation}
-x_0^2+x_1^2+x_2^2+ \cdots + x_d^2=-R_h^2 \ ,
\label{hype}
\end{equation}
embedded in $(d+1)$-dimensional Minkowski space. In the simple case $d=3$, of
particular interest in this paper, we can use the coordinate identifications
\begin{eqnarray}\label{coord}
 x_0=R_h \cosh\chi\, , &\quad & x_1=R_h \sinh\chi \cos\alpha\, , \nonumber \\
 x_2=R_h \sinh\chi\sin\alpha\cos\beta\, , &\quad & x_3=R_h \sinh\chi\sin\alpha\sin\beta
\, ,
\end{eqnarray}
to relate this representation to the induced metric
\begin{equation}
ds^2=R_h^2 \left[d\chi^2 +\sinh^2(\chi)\left(d\alpha^2 + \sin^2(\beta)d\beta^2 \right)\right]
\end{equation}
on $H^3$.
>From this perspective it is
easy to understand why the isometries of $H^3$ are described by the
orientation preserving homogeneous Lorentz group in
4-dimensions, $SO(3,1)$.

\begin{figure}
\centerline{\epsfbox{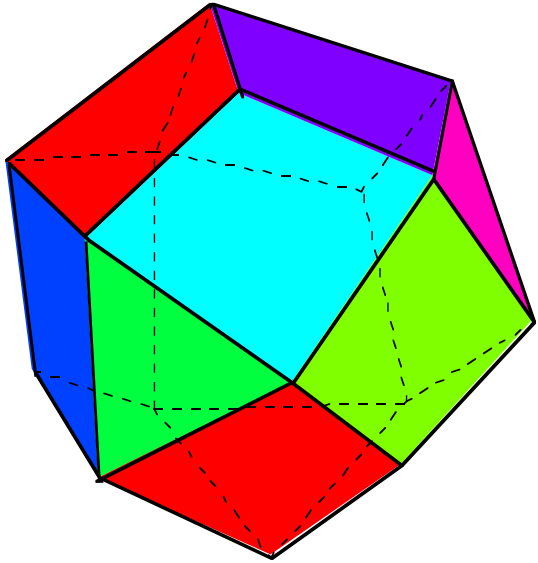}}
\caption{The Thurston Manifold.}
\label{thurston}
\end{figure}

To illustrate the features of compact hyperbolic spaces, we will consider the
Thurston manifold \cite{thurston} (see Fig.~\ref{thurston}).
One particularly useful way to study this and other compact
hyperbolic spaces is to use the {\it SnapPea} \cite{weeks} catalogue. The Thurston
manifold $\Sigma_{{\rm Th}}$ (m003(-2,3) in the {\it SnapPea} census) has fundamental
group, $\Gamma=\pi_1(\Sigma_{{\rm Th}})$, with presentation
\begin{equation}\label{fun}
\Gamma=\{ a,b \, : \, a^2ba^{-1}b^3a^{-1}b,\;
ababa^{-1}b^{-1}ab^{-1}a^{-1}b \} \, .
\end{equation}

Here $a$ and $b$ are the generators of the fundamental group, describing
identifications in
the faces of the fundamental cell shown in Fig.~\ref{thurston}, and in usual group-theoretic notation
the expressions following the colon in equation~(\ref{fun}) are
set equal to the identify.

The fundamental
cell is drawn using Klein's projective model for hyperbolic space. In
this projection $H^3$ is mapped into an open ball in Euclidean $3$-space $E^3$.
Under this
mapping hyperbolic lines and planes are mapped into their Euclidean
counterparts. This is why the totally geodesic faces of the
fundamental cell appear as flat planes.
Thurston's manifold
has volume approximately $0.98R_h^3$.

By acting on points lying on the symmetry axis of each group element
it is possible to compile a list of the the minimal geodesics. A
typical isometry is a Clifford translation -- a corkscrew type motion, consisting
of a translation of length $L$ along a geodesic, combined with
a simultaneous rotation through an angle $\omega$ about the
same geodesic. The length and torsion can be found directly from
the eigenvalues of the group element, and are conveniently listed
by the {\em SnapPea} program\cite{weeks}.

We can make some general observations about the existence of long wavelength
modes on $H^3/\Gamma$. In large volume CHM's there is
generically a gap in the spectrum of the Laplacian (more specifically,
the Laplace-Beltrami operator) between the
zero mode  and the next lowest mode. A theorem due
to Sarnak in $d=2$ and a conjecture due to Brooks in $d\geq3$
state (approximately) that for large volumes characteristically
$m_{\rm gap} = {\cal O}(R_h^{-1})$. This puts an upper limit on the wavelength
of modes.

We are interested in CHMs as extra-dimensional manifolds.
Because they are locally negatively curved, CHM's exist only for $d\ge 2$.
Their properties are well understood only for $d\leq3$, however, it is known
that CHM's in dimensions $d\geq 3$ possess the important property of
{\em rigidity} \cite{MostowPrasad}.
As a result, these manifolds have no massless shape moduli.
Hence, the stabilization of such internal
spaces reduces to the problem of stabilizing a single modulus,
the curvature length or the ``radion.''

The primary reason for considering such manifolds for compactification is the
behavior of their volume as a function of linear size.
In general, the total volume of a smooth compact hyperbolic space in any number
of dimensions is
\begin{equation}
{\rm Vol}_{\rm new}=R_h^d~ e^{\alpha } \ ,
\label{volume}
\end{equation}
where $R_h$ is the curvature radius and $\alpha$ is a constant
{\em determined by topology}.
(For $d=3$ it is known that there is a countable infinity of
orientable CHM's, with dimensionless volumes, $e^\alpha$,
bounded from below, but unbounded from above; moreover
the $e^\alpha$ do not become sparsely distributed with large volume.)
In addition, because the topological invariant $e^{\alpha}$ characterizes
the volume of the CHM,
it is also a measure of the largest distance $L$ around the manifold.
CHM's are globally anisotropic; however,
since the largest linear dimension gives the most significant contribution to
the volume, there exists an approximate relationship between
$L$ and ${\rm Vol}_{\rm new}$.
For $L \gg R_h/2$ the appropriate asymptotic relation,
dropping irrelevant angular factors, is
\begin{equation}
\alpha \simeq \frac{(d_{eff}-1)L}{R_h} \ ,
\end{equation}
where $1< d_{eff}\leq d$.

Thus, in strong contrast to the flat case, the expression for
$M_P$ depends {\em exponentially on the linear size},
\begin{equation}
M_P^2=M_{d+4}^{2+d}R_h^d e^{\alpha } \simeq M_{d+4}^{2+d} R^d_c
\exp\left[\frac{(d_{eff}-1)L}{R_h}\right] \ .
\end{equation}
The most interesting case (and as we will see later, most
reasonable) is the smallest possible curvature radius,
$R_h \sim M_{d+4}^{-1}$.  Taking $ M_{d+4}\sim $~TeV then yields
(with $d_{eff}=d=3$)
\begin{equation}
\label{lmaxnum}
L\simeq 35 M_{d+4}^{-1} = 10^{-15} {\rm mm} \, .
\end{equation}
Therefore, one of the
most attractive features of CHM's is that to generate an exponential
hierarchy between $M_{d+4}\sim$~TeV, and $M_P$ only requires that the linear size
$L$ be very mildly tuned if the internal space is a CHM.

\section{Hyperbolic Brane World Cosmology}
\label{stabilization}
Our starting point is the action for Einstein gravity in a $4+d+n$-dimensional
space-time, with bulk matter.
\bea
S=\int d^{4+d+n}x \sqrt{-G}\left[M^{d+n+2}R(G)-{\cal L}_{\rm bulk}\right] \ ,
\eea
where $M$ is the $4+d+n$-dimensional Planck mass and ${\cal L}_{\rm bulk}$ is the
Lagrangian density. We will assume that the geometry of the bulk space $\Sigma^{4+d+n}$
is factorizable into the form
\be
\Sigma^{4+d+n}={\cal F}^{3+1} \times H^d/\Gamma \times S^n \ ,
\ee
where ${\cal F}^{3+1}$ denotes a $3+1$-dimensional Friedmann, Robertson-Walker (FRW) space,
$H^d/\Gamma$ is a $d$-dimensional compact hyperbolic manifold and $S^n$ is the
$n$-sphere, with volume $\Omega_n$.

We have included a spherical factor here because of the hope that its curvature will play a role
in cancelling that of the hyperboloid. As we shall see this is not sufficient. In demonstrating
this it will become clear that adding other factors will not help the situation. Therefore this
choice of manifold seems sufficiently general to prove our result.

The metric ansatz consistent with this factorization is
\be
ds^2=G_{AB}dx^Adx^B=\bar{g}_{\mu\nu}dx^\mu dx^\nu +
r_h^2\gamma_{ij}dy^idy^j +r_s^2\omega_{ab}dz^a dz^b\ .
\ee
Here $A,B,\ldots =0\ldots 3+d+n$ are indices on the whole bulk space-time,
$\mu ,\nu ,\ldots = 0\ldots 3$ are indices on the $3+1$ dimensional brane,
$i,j,\ldots = 4\ldots d+3$
are indices on the CHM and $a,b,\ldots = d+4\ldots d+n+3$ are indices on the
sphere. The metric on the brane is denoted by $\bar{g}_{\mu\nu}$, that on the unit $d$-hyperboloid
by $\gamma_{ij}$ and the metric on the unit $n$-sphere by $\omega_{ab}$. There are therefore
two radion fields in the problem, $r_h$, the curvature radius of the CHM, and $r_s$, the
curvature radius of the sphere. We will denote the values of these radii at their
putative stable points by $R_h$ and $R_s$ respectively. By the volume considerations
of the previous section, the effective $3+1$-dimensional Planck mass will then be given by
\be
M^2_4=M^{2+d+n}(R_h^d e^\alpha)(R_s^n \Omega_n) \ .
\ee
There are two ways to analyze the issue of stabilization in extra-dimensional theories.
We may consider the full $4+d+n$-dimensional equations, or those of the dimensionally
reduced $3+1$ dimensional theory. For completeness we will express the problem in an
effective theory setting and then demonstrate our main result in the full theory.

To derive our effective theory, let us define the fields $\phi$ and $\psi$ by
\bea
r_h &=& R_h \exp{\left[\sqrt{\frac{1}{d(d+2)}}\frac{\phi}{M_4}\right]} \nonumber \\
r_s &=& R_s \exp{\left[\sqrt{\frac{1}{n(n+2)}}\frac{\psi}{M_4}\right]} \ ,
\eea
and perform a conformal rescaling of the brane metric
\be
\bar{g}_{\mu\nu}=g_{\mu\nu} \exp{\left[-\sqrt{\frac{d}{d+2}}\frac{\phi}{M_4}-\sqrt{\frac{n}{n+2}}
\frac{\psi}{M_4}\right]} \ .
\ee
This decouples $\phi$ and $\psi$ from the reduced Einstein tensor.
Integrating over the compact manifolds, we may now define an effective action $S_{\rm eff}$ by
\bea
S_{4+d+n}=\int d^{d+n}x{\sqrt{\gamma}\over e^\alpha}{\sqrt{\omega} \over \Omega_n}S_{\rm eff} \ ,
\eea
with
\bea
S_{\rm eff}=\int d^4x \sqrt{-g}\left[M^2_4R(g) - {1\over 2}(\nabla\phi)^2
-{1\over 2}(\nabla\psi)^2 \right.\nonumber \\
\left. - 2 \sqrt{{nd\over(d+2)(n+2)}}\nabla\phi\nabla\psi -W(\phi,\psi,g)\right] \ .
\eea
Here, the effective rescaled potential is
\bea
W(\phi,\psi,g)&=&\left(\frac{M_4^2}{M^{n+d+2}}\right){\cal L}_{bulk}
\exp\left[-\sqrt{\frac{d}{d+2}}\left(\frac{\phi}{M_4}\right)-
\sqrt{\frac{n}{n+2}}\left(\frac{\psi}{M_4}\right)\right]+
\nonumber \\
&+& \left[\frac{d(d-1)}{R_h^2}-\frac{n(n-1)}{R_s^2}\right]M_4^2
\exp\left[-\sqrt{\frac{d+2}{d}}\left(\frac{\phi}{M_4}\right)-
\sqrt{\frac{n+2}{n}}\left(\frac{\psi}{M_4}\right)\right] \ ,
\eea
where we have used that $R(\omega)=n(n-1)$.

In this language the stabilization of the two
radii translates into the following obvious system of equations:
\bea
\partial_\phi W|_{(\phi,\psi)=0}=0\;\;\; & , & \;\;\;\partial^2_\phi
W|_{(\phi,\psi)=0}>0 \nonumber \\
\partial_\psi W|_{(\phi,\psi)=0}=0\;\;\; & , & \;\;\;\partial^2_\psi
W|_{(\phi,\psi)=0}>0 \ ,
\label{condition1}
\eea
plus the condition that the effective four-dimensional cosmological constant vanish,
\be
\frac{\delta S_{\rm eff}}{\delta g^{\mu\nu}}=\left.\left(g_{\mu\nu}W -2\frac{\partial W}{\partial
g^{\mu\nu}} \right)\right|_{(\phi,\psi)=0}=0 \ .
\label{condition2}
\ee

In the full theory we shall adopt a cosmological approach and assume that, whatever bulk
matter is present, its energy-momentum tensor can be expressed in perfect fluid form on each
of the submanifolds
\bea
T_{00} &=& \rho \nonumber \\
T_{\alpha\beta} &=& p{\bar g}_{\alpha\beta} \nonumber \\
T_{ij} &=& qr_h^2\gamma_{ij} \nonumber \\
T_{ab} &=& sr_s^2\omega_{ab} \ ,
\label{emtensor}
\eea
where $\alpha$, $\beta =1..3$. The Einstein equations
then become
\bea
&&\rho= 3\left[{k\over a^2}+\left({\dot{a}\over a}\right)^2\right]
+{1\over 2}\left[\frac{n(n-1)}{R^2_s}-\frac{d(d-1)}{R^2_h}\right]M^{2+d+n} \nonumber\\
&&p=-\left[{k\over a^2}+\left({\dot{a}\over a}\right)^2 + 2{\ddot{a}\over a}\right]
-{1\over 2}\left[\frac{n(n-1)}{R^2_s}-\frac{d(d-1)}{R^2_h}\right]M^{2+d+n}\nonumber\\
&&q=-3\left[{k\over a^2}+\left({\dot{a}\over a}\right)^2+{\ddot{a}\over a}\right]
-{1\over 2}\left[\frac{2n(n-1)}{R^2_s}-\frac{(d-1)(d-2)}{R^2_h}\right]M^{2+d+n}\nonumber \\
&&s=-3\left[{k\over a^2}+\left({\dot{a}\over a}\right)^2+{\ddot{a}\over a}\right]
-{1\over 2}\left[\frac{(n-1)(n-2)}{R^2_s}-\frac{2d(d-1)}{R^2_h}\right]M^{2+d+n} \ .
\label{eqn5}
\eea
The null energy condition, $T_{AB}N^A N^B\geq 0$ for all null $4+d+n$-vectors $N^A$, in
conjunction with the ansatz~(\ref{emtensor}) yields
\bea
\rho + p &\geq & 0 \nonumber \\
\rho + q &\geq & 0 \nonumber \\
\rho + s &\geq & 0 \ ,
\eea
which then implies
\bea
{k\over a^2}+\left({\dot{a}\over a}\right)^2 -{\ddot{a}\over a} &\geq& 0 \label{nec1} \\
{\ddot{a}\over a} &\leq& -{1\over 6}\left[\frac{n(n-1)}{R^2_s}+\frac{2(d-1)}{R^2_h}\right] \label{nec2}\\
{\ddot{a}\over a} &\leq& {1\over 6}\left[\frac{2(n-1)}{R^2_s}+\frac{d(d-1)}{R^2_h}\right]  \ .
\label{nec3}
\eea

It is relatively straightforward to see how these inequalities are incompatible with a
reasonable cosmological evolution on the brane. Note first that~(\ref{nec2}) implies~(\ref{nec3}).
Thus~(\ref{nec2}) is the important inequality to deal with. Successful $3+1$-dimensional
cosmology requires a radiation dominated phase (in order that successful nucleosynthesis occur),
followed by a matter dominated phase. Focusing on the radiation dominated phase, the
scale factor evolves as $a(t)\propto t^{1/2}$, which means that
\be
\left.\frac{{\ddot a}}{a}\right|_{\rm radiation} = -\frac{1}{4}\frac{1}{t^2} \ .
\ee
Therefore, certainly when the universe is older than than $R_h^{-1} \sim ({\rm TeV})^{-1}$, the
null energy condition is violated. Since nucleosynthesis occurs at a later time than this, it
is clear that, even in the most optimistic case the model is not cosmologically viable.

Therefore we conclude that any bulk matter that stabilizes the radion and gives rise to an
acceptable cosmology on the brane must violate the null energy condition.

Note further that, in the special case that we restrict our brane to be $3+1$-dimensional
Minkowski space-time and restrict the extra-dimensional manifold to be purely a CHM (no $S^n$
factor),  our constraints simplify (see \cite{Carroll:2001ih}) considerably to become
\bea
\rho &=& -\frac{d(d-1)M^{2+d}}{2R^2_h} \nonumber\\
p &=& \frac{d(d-1)M^{2+d}}{2R^2_h}\nonumber\\
q &=& \frac{(d-1)(d-2)M^{2+d}}{4R^2_h} \ .
\label{eqn3}
\eea
In this case it is simple to see that the required matter field cannot obey the null
energy condition. Consider a general null vector with spatial components in the
direction of the hyperboloid, for example
\be
N = \partial_t+e_i \ ,
\ee
where $\{e_i\}$ is an orthogonal vector basis on the hyperboloid. Contracting the energy-momentum
tensor twice with this vector yields
\be
T_{AB}N^A N^B = -\frac{(d-1)(d+2)M^{2+d}}{4R^2_h} <0 \ .
\ee
Hence the required matter must violate the null energy condition.

To clarify the above general statement, let us consider a simple
example of bulk matter that one might expect to stabilize our CHM.
For this example we'll stay within the simplified context of requiring Minkowski space
to be the solution on our brane, as above. Our bulk matter will consist of a cosmological
constant $\Lambda=M^{d+4}\lambda$ and a d-form $F_{[d]}$ over the hyperbolic manifold.
The bulk Lagrangian density is therefore
\be
{\cal L}_{\rm bulk}=\Lambda+\frac{F_{[d]}^2}{2d!} \ .
\ee
The resulting field equations for $F_{[d]}$ can be solved by the ansatz
\bea
F_{\mu ABC} &=& 0 \nonumber \\
F_{45..d+4} &=& B(x^{\mu}) \ ,
\eea
so that $B$ is independent of the hyperbolic coordinates. With this ansatz there results a
trapped magnetic flux on the compact hyperbolic space and the constant $B$ is related to the
radion $r_h$ by the equation
\be
B=bM^{(d+4)/2}\left(\frac{R_h}{r_h}\right)^d \ .
\ee
Thus the on-shell form of the bulk Lagrangian density becomes
\be
{\cal L}_{\rm bulk}=M^{d+4}\left[\Lambda+{b^2\over 2}\left({R_h\over r_h}\right)^{2d}\right] \ .
\ee
Defining $\beta^{-1}=MR_h$, equations~(\ref{condition1}) and~(\ref{condition2}) reduce to
\bea
\Lambda +{b^2 \over 2} +d(d-1)\beta^2 &=& 0, \nonumber \\
b^2d+d(d-1)(d+2)\beta^2 &=& 0 \ .
\eea
The corresponding solution is
\bea
\Lambda &=& -(d-1)(d-2)\frac{\beta^2}{2} \nonumber \\
b^2 &=& -(d-1)(d+2)\beta^2 \ .
\eea
Using these solutions, the on-shell stress energy tensor of
the d-form becomes
\be
T_{AB}={(d-1)(d-2)\beta^2\over
8}(g_{AB}-\delta^{ij}_{AB}g_{ij}) \ ,
\ee
therefore when contracted
twice with a null vector along one of the hyperbolic directions
(i.e. $N=N^0e_0+N^ie_i$), we obtain
\be
T_{AB} N^A N^B=-{(d-1)(d-2)\beta^2\over 8}g_{ii}N^i N^i <0 \ ,
\ee
clearly violating the null energy condition.

\section{Going Beyond Classical Matter : The Casimir Force}

Although classical fluids are expected to respect the null energy condition,
it is well known that such conditions are violated quantum mechanically.  A specific
example of such violation is Casimir energy density -- the zero point energy
density of the quantum fields on the space time.  The magnitude and sign of the
Casimir energy depend in complicated ways on the topology and geometry of the
underlying manifold, and on the specific field content (for an example of the use
of the Casimir force in stabilizing extra-dimensional models see \cite{Ponton:2001hq}).

Let us continue to consider the simplified example space-time and further focus on the
case in which the extra-dimensional manifold is $3$-dimensional,
\begin{equation}
M = M^{(4)} \times H^3/\Gamma \ .
\end{equation}
The Casimir energy density will depend on both the details of $\Gamma$,
and on the curvature $R_h$ of the $H^3$,
\begin{equation}
\rho_C \propto R_h^{-p} \ ,
\end{equation}
with $p>0$. Thus the Casimir pressure in the $H^3/\Gamma$ directions will be proportional
to $R_h^{-p-1}$, while the pressure will be zero in the Minkowski space directions.
Since the null energy
condition would require that when the Casimir energy density, $\rho_C$ is negative
the associated pressure is isotropic and equal to $-\rho_C$,
the null energy condition is therefore violated. Interestingly, with $\rho_C<0$,
the sign of the  energy density and the pressure are precisely right to help
stabilize $H^3/\Gamma$.

Because of the chaotic nature of flows on CHMs, we do not present any generic results
for the value of $\rho_C$ in specific CHMs, except to note that the case
of a minimally coupled scalar field on the Thurston manifold has been studied numerically
by Miller, Fagundes and Opher \cite{Muller:2001pe}.  In this study they discovered that the Casimir
energy density can be negative, as needed for stabilizing the extra-dimensional manifold in our
model. An understanding of the Casimir energy in the specific highly topologically complex
manifolds used in our model would be extremely useful, but
seems calculationally formidable and is beyond the scope of this paper.

\section{Supersymmetry in Hyperbolic Backgrounds}
A question of great interest when suggesting any compactification scheme is that
of low-energy supersymmetry. For the case of interest in this paper, we may begin with an
explicit construction of the Killing spinors of maximally
symmetric spaces with negative cosmological constant \cite{lpt1,lpr1}.
For the space $H^d$ we choose coordinates in the horospherical frame in which
the metric takes the form,
\be
ds^2=e^{2r}\delta_{\alpha\beta}dx^\alpha dx^\beta+dr^2 \ .
\ee
In this frame, the Killing spinors are given by,
\be
\xi = e^{{1\over2} r\Gamma_r}\left[1+{1\over2}
x^\alpha\Gamma_\alpha(1-\Gamma_r)\right]\epsilon \ ,
\ee
where $\epsilon$ is an arbitrary constant spinor (the cases d=2,3 are
special and expressions can be found in ref~ \cite{fy1}).

We can see from this expression that the number of supersymmetries of $H^d$ is equal to
the number of independent spinor components. Now, recall that the isometry group
is $SO(1,d)$ and that compact hyperbolic
manifolds are obtained by quotient of $H^d$ by a discrete subgroup
$\Gamma$ of $SO(1,d)$, with no fixed points. Whether or not any Killing spinors
survive this quotienting process depends on $\Gamma$ \cite{Kehagias:2000dg}.
\begin{enumerate}
\item If d is
even then the spinors are in an $SO(1,d-1)$ representation, and all supersymmetries are broken.
\item If $d$ is odd, then the spinors are in an $SO(1,d)$ representation. In this case
there are several possibilities.
\begin{enumerate}
\item If $\Gamma$ is a subgroup of $SO(1,d-1)$ some Killing spinors may survive, since
we can decompose the original Killing spinors into Weyl spinors on
the representation of this group.
\item If $\Gamma$ is a not subgroup of $SO(1,d-1)$ then all supersymmetries are broken.
\item In the special case $d=3$ there also remain no supersymmetries.
\end{enumerate}
\end{enumerate}

\section{Conclusions}
The initial motivation for the idea of large extra
dimensions was to address the hierarchy problem. Allowing the internal space to be a compact
hyperbolic manifold \cite{Kaloper:2000jb} introduces the attractive new feature that the
hierarchy problem becomes a truly mild tuning of the length scales in the theory.
In addition, such models have been shown to have interesting
cosmological consequences \cite{Starkman:2001dy,Starkman:2001xu}.
Of course, in all extra-dimensional models, the stabilization
of the radion mode is an essential component, since the effective four-dimensional
gravitational constant does not change today. One advantage of CHMs as internal spaces is that
in three or more dimensions the property of rigidity means that there is a single radion --
the curvature radius of the manifold. Nevertheless, stabilization of this single mode remains
an important issue in these models.

In this paper we have analyzed the constraints on the
bulk matter that is required to achieve this necessary stabilization. We have allowed for the
metric on the brane to be cosmological, of FRW form, and in this sense we have expanded on the
elegant results of \cite{Carroll:2001ih} which apply to the case that the brane is Minkowski.
Our results apply to extra dimensions large enough that general relativity is a valid
theory, and show that stabilization may only be achieved by bulk matter that violates the
weak energy condition.  We have demonstrated this result explicitly, and have provided an example of
how a specific example of matter violates this condition.

Since such violations are problematic for classical matter,
quantum effects may be crucial in stabilizing these manifolds. In fact, since the essential
feature that makes CHMs so attractive is their large volume without large linear spatial extent,
the assumption that general relativity is completely valid may break down. As a specific
example we have suggested the Casimir effect as such a quantum contribution.
While simple estimates indicate
that this may be a successful way to stabilize CHMs, we caution that to understand the effect
in the manifolds that are useful for brane-world models is a much more difficult task.

Finally,although independent from the issue of stabilization, for completeness we have briefly
discussed the ways in which supersymmetry in compact hyperbolic extra dimensions can be broken
due to the obstructions to defining covariantly constant spinors on these spaces.

\vspace{1cm}
\noindent
{\bf Acknowledgments}\\
The work of SN and GDS is supported by grants from the U.S. Department of Energy. The work of
PS is supported in part by the National Science Foundation (NSF) under grant PHY-0098747 and by
funds from Syracuse University. The work of MT is supported by the NSF under grant PHY-0094122.
GDS thanks Nemanja Kaloper  for extensive discussions of the Casimir energy density and thanks the
Aspen Center for Physics, where some of this work was completed. MT thanks Sean Carroll and
Mark Hoffman for useful discussions.


\end{document}